\documentstyle[12pt]{article}
\begin{document}
\rightline {DFTUZ 92/25}
\vskip 1.0cm
{\large\centerline{\bf Phase Structure of Compact Lattice $QED_{3}$}
\vskip 0.3cm
\centerline{\bf with Massless Fermions}}
\vskip 1.0cm
\centerline{Vicente ~Azcoiti and Xiang-Qian ~Luo}
\vskip 0.6cm
\centerline{\it Departamento de F\'\i sica Te\'orica, Facultad de
Ciencias,}
\centerline{\it 50009 Zaragoza, Spain}

\vskip 4.0cm

\centerline{\bf Abstract}
\vskip 0.3cm
\par
{In the framework of (2+1)-dimensional compact lattice QED with
light fermions, we
investigate the phase diagram in the $(\beta, N)$ plane. The approximations
involved are related to an expansion of the effective fermionic action as a
power series of the flavor number $N$. We also develop a new
mechanism for understanding the $N-$critical phenomenon in the full theory.
Our results for the specific heat
indicate that only one phase does exist.
We give strong evidences that this qualitative result should not be
changed with the inclusion of higher order terms in the $N$ expansion.}

\eject

%%%%%%%%%%%%%%%%%%%%%%%%%%%

There are several reasons which made it interesting the study
of Quantuum Field Theories in 2+1 dimensions and justify the
increasing attention devoted to this field in recent time.
Essentially, the motivation for such an analysis is two-fold:

i) There are many similarities between gauge theories in 2+1
dimensions and four dimensional QCD like quark confinement and
spontaneous symmetry breaking, which suggest us to use three
dimensional formulation of gauge theories as a simpler laboratory
to study general features of QCD.

ii) Recently it has been shown that a very popular model in the
area of high $T_c$ superconductivity, the Heisenberg model, can
be reformulated exactly as a $SU(2)$ lattice gauge theory with
massless dynamical staggered fermions in 2+1 dimensions at strong
coupling ($\beta=0$) [1]. Chiral-symmetry breaking in
three-dimensional lattice gauge theories is the analog of the Neel
phase of antiferromagnets. On the other hand the fact that we do not
expect essential differences between the abelian and non abelian
models, at least in the strong coupling region, suggests that the
study of the abelian case should be relevant also in this context.

The first analytic studies of continuum $QED_{3}$ in the
$1/N$ approximation [2,3] indicated that dynamical fermion mass
generation takes place for any arbitrarily large number of flavors
$N$. More recently, a reexamination of the $1/N$ expanssion of the
Dyson-Schwinger kernel opened the possibility for the existence of
a critical $N_c \approx 32/ \pi^2$ beyond which chiral symmetry
is restored [4,5].

The results of the Dyson-Schwinger approach [4,5] seem to have been
confirmed by the first numerical simulations of noncompact $QED_3$
[6,7]. However the small lattices and limited data for the fermion
masses $m$ used in [6,7] to extrapolate to the chiral limit, makes
this results not conclusive. Furthermore Pennington and Walsh [8]
have recently found by numerical methods non perturbative solutions
of the Dyson-Schwinger equation for the full fermion propagator,
which exhibit chiral symmetry breaking for all values of $N$.
The corresponding chiral condensate seems to decrease exponentially
with $N$ and if it is true, this kind of behavior puts serious doubts
on the feasibility of the $1/N$  approximation approach to $QED_3$.

Therefore, at present the situation is far from being clear and
the aim of this letter is to increase our understanding of the
$N$-critical phenomenon by studying numerically the three-dimensional
compact model.
We will pay special attention to the analysis of
the phase diagram of this model in the $(\beta, N)$ plane and will
develope a physical picture for understanding the origin of
phase transitions in this parameter space. A similar study to that
developed here  but for the noncompact model is in progress and the
results will be published in the next future.

The approach we have used in our calculations is based on the application
of the microcanonical fermionic average method introduced in [9] and
tested for the compact [10] and noncompact [11,12] formulations of the
abelian model in four dimensions. We therefore refer to these references
for technical details.

In order to explore the phase diagram in the $(\beta, N)$ plane, we
should search for non analyticities of thermodynamical quantities in
this plane. A direct investigation of the zero mass limit of the chiral
condensate, which would give us information on how chiral symmetry is
realised in this model at different flavor numbers, has several technical
difficulties. Mainly, an extrapolation of the numerical results for the
chiral condensate to the massless limit is necessary and the final result
may depend on the kind of fitting function used in this procedure,
specially when the numerical value of the chiral condensate is very small.

This difficulty can be overcome if we look for other thermodynamical
quantities like plaquette energy and specific heat since in such a
case no symmetry forces them to be zero. This is indeed what was done
in [11,12] to investigate the phase diagram and the nature of the fixed
point in (3+1)-dimensional noncompact QED and the results were very
satisfactory. Therefore in this letter we will concentrate on the
investigation of the behavior of the plaquette energy and specific heat
in the $(\beta, N)$ plane whereas our results for the chiral condensate
will be reported in a separate publication.

Our starting point is the definition of an effective fermionic action [9]
which depends on the pure gauge plaquette energy E, fermion mass m and
flavor number $N$ and which is related to the gauge link variables
by the expression

\begin{equation}
e^{-S_{eff}^F(E,m,N)}=
{\int [dU_{\mu}(x)] (det
\Delta(m, U_{\mu}(x)))^{N/2}
\delta(S_G (U_{\mu}(x)) - 3VE)
\over
\int [dU_{\mu}(x)]
\delta(S_G (U_{\mu}(x)) - 3VE)},
\end{equation}

\noindent
where the integration is over all gauge link variables that are
elements of the compact $U(1)$ gauge group. V is the lattice volume,
$\Delta$ is the fermionic matrix (we use staggered fermions) and the
exponent $1/2$ in the determinant is due to the fermion doubling.
$S_G(U)$ is the full pure gauge plaquette energy defined as the sum
over all lattice plaquettes of the cosine of the plaquette angle and
the denominator in (1) is the density of states $N(E)$ of fixed pure
gauge normalized energy $E$.

The partition function can be written now as a one-dimensional integral
over
the normalized pure gauge energy as

\begin{equation}
{\cal Z} = \int dE e^{-S_{eff}(E,m,N,{\beta})}.
\end{equation}

\noindent
where $S_{eff}(E,m,N,{\beta})$ is the full effective action related to
the density of states $N(E)$, the fermionic action $S^{F}_{eff}(E,m,N)$
and the pure gauge energy $E$ by

\begin{equation}
S_{eff}(E,m,N,{\beta})= - ln N(E) - 3 \beta VE + S^{F}_{eff}(E,m,N).
\end{equation}

\noindent
This effective action diverges linearly with the lattice volume $V$ in the
infinite volume limit and in this limit, the thermodynamics of the system
can be studied by means of the saddle point technique. The mean plaquette
energy $<E_p> = E_0(m, \beta, N)$ will be given by the solution of the
saddle point equation

\begin{equation}
{-1 \over n(E)} {dn(E) \over dE}+
{\partial \overline S^{F}_{eff}(E,m,N) \over
\partial E} - 3 \beta = 0
\end{equation}

\noindent
satisfying the minimun condition

\begin{equation}
\lbrace {1 \over n(E)^{2}} {dn(E) \over dE}^{2} -
{1 \over n(E)}{d^{2}n(E) \over dE^{2}}
+ {\partial^{2} \overline S^{F}_{eff}(E,m,N) \over
\partial E^{2}} \rbrace_{E_{0}(m,\beta,N)} > 0,
\end{equation}

\noindent
where $n(E)$ in (4), (5) is related to the density of states $N(E)$ by the
expression

\begin{equation}
n(E) = N(E)^{1/V},
\end{equation}

\noindent
and $\overline S_{eff}^F(E,m,N)$ is the effective fermionic action
normalized by the lattice volume.

The specific heat $C_{\beta} = {{\partial E_0}/ \partial {\beta}}$ can
be obtained by derivating equation (4), and the final expression is

\begin{equation}
C_{\beta} = 3 \lbrace {1 \over n(E)^{2}} {dn(E) \over dE}^{2} -
{1 \over n(E)}{d^{2}n(E) \over dE^{2}}
+ {\partial^{2} \overline S^{F}_{eff}(E,m,N) \over
\partial E^{2}}\rbrace_{E_{0}(m,\beta,N)}^{-1}.
\end{equation}

Before discussing the results of our numerical simulations, let us do a
qualitative analysis on the basis of
expression (7) for the specific heat. The
question now is: how a phase transition can be generated when the number of
dynamical flavors is large enough? In the pure gauge model it has been shown
that no phase transition occurs at any finite value of the inverse
coupling constant $\beta$ [13,14]. In other words moving $\beta$ from zero to
$\infty$, the plaquette energy takes continuously all the values from 0 to 1.
Therefore the sum of the first two terms in the denominator of the specific
heat (7) will be positive in all the [0,1) energy interval (convexity of the
effective action!). Now let switch on dynamical fermions in the system and
consider the $N$ parameter as a continuous
parameter. It can be shown that the
effective fermionic action $\overline S_{eff}^F (E,m,N)$ diverges linearly
with $N$ when
N goes to $\infty$. Then, if the sign of the second energy derivative of
$\overline S_{eff}^F (E,m,N)$ is negative in some energy interval
and $N$ is large
enough, the third contribution to the denominator of the specific heat (7)
can compensate the pure gauge contribution and the global sign in the
denominator of (7) could be changed in this way [15]. In such a case there will
be some energy interval in which the minimun condition (5) of the saddle point
equation (4) does not hold. This energy interval will not be accesible to
the system and then a first order phase transition will appear. Decreasing
now the value of the parameter $N$, we will find a critical value $N_c$, at
which the energy interval where equation (5) does not hold will
become a single point. This critical value $N_c$, where the specific heat
diverges, will be the end point of a first order phase transition line in the
$(\beta, N)$ plane.

Now let us  analyse our numerical results in order to see how this general
scheme is realized in compact $QED_3$. The feasibility of a direct computation
of the effective fermionic action (1) is related to the form of the probability
distribution function of the logarithm of the fermionic determinant at fixed
pure gauge energy [16]. An alternative and more reliable way is to compute
$S_{eff}^F$ by expanding it in a power series of $N$
(cumulant expansion) [12,16]

\begin{equation}
-S_{eff}^F (E,m,N) = {N\over 2} < ln det \Delta>_E
+ {N^2 \over 8}
\lbrace <(ln det \Delta)^2>_E - <ln det \Delta >_{E}^2 \rbrace + ...
\end{equation}

\noindent
where $<O>_E$ means the mean value of the operator $O(U_{\mu}(x))$ computed
with the probability distribution $[dU_{\mu}(x)]
\delta(S_G (U_{\mu}(x) - 3VE)$.

We have done simulations on $8^3$, $10^3$ and $14^3$ lattices. On each lattice
$100000$ configurations at each fixed $E$ were generated by an exact
microcanonical algorithm. The fermionic matrix at zero mass was exactly
diagonalized for 200 decorrelated configurations at each energy by means of a
modified Lanczos algorithm [17]. From the eigenvalues obtained in this way
we can compute the succesive terms in (8) for any value of the fermion mass
$m$. Fig. 1 shows the $<ln det\Delta>_E$ as a function of E
on a $14^3$ lattice.

We have done canonical simulations of the effective theory described by the
partition function (2). In these simulations we have approximated the effective
fermionic action by its first contribution in the cumulant expansion (8) and
measured the plaquette energy, specific heat, chiral condensate and Binder
parameter. Details of these simulations will be reported in a longer
publication.
The most relevant thing which emerges from the analysis of the results of
these simulations is the ($\beta, N$) plane phase diagram
for massless fermions plotted
in Fig. 2. The striking feature of this phase diagram is the fact that the
first order phase transition line ends at finite (negative) $\beta$.

As a check of these results we have done the following simple calculation.
We fitted the results for the first contribution to the effective fermionic
action by a fifth degree polynomial (continuous line in Fig. 1). From the
results of this fit and the computation of the specific heat in the pure
gauge model, we found by means of equation (5) that the critical number of
flavors at the end point of the first order phase transition line is
$N_c = 18.4$ and the critical plaquette energy $E_c = 0.48$, in very good
agreement with the results of the numerical simulation.

The fact that the first order phase transition line of Fig. 2 ends at a
finite value of the inverse coupling constant $\beta$ can be understood on
the light of the energy dependence of the effective fermionic action plotted
in Fig. 1. The second energy derivative of the effective fermionic action is
negative in the small-intermediate energy region but it becames positive in
the large energy region. Therefore at large energy (or equivalently large
$\beta$ values) the specific heat will be finite and equation (5) will hold.
In order to get a phase transition line ending at $\beta = \infty$
an opposite convexity of $S_{eff}^F$ in the large energy region is necessary.

Might this qualitative result be changed with the inclusion of higher order
terms
in the $N$ expansion of the effective fermionic action? We think that there
are several indications which make such a possibility highly
unlikely. They are
mainly:

i) The second energy derivative of the pure gauge effective action (sum of the
first two terms in (5)) is, as stated before, positive for all $E>0$. Even
more, when the energy approaches the maximum value $E=1$, this quantity
diverges or equivalently the specific heat goes to zero when $\beta$ goes to
$\infty$ in the pure gauge theory. Therefore in order to change the sign of
(5) at $E$ near 1, we need not only a change in the convexity of the effective
fermionic action at large energy but also a second energy derivative negatively
divergent in this limit.

ii) The probability distribution of the logarithm of the fermionic determinant
at large energy is very narrow (much more than in the small energy region) due
to the fact that the density of states $n(E)$ goes to zero when $E$ goes to 1.
In fact we have observed that the fluctuations of the logarithm of the
determinant at fixed energy in the large energy region $(E>0.85)$ are
indipendent of the lattice volume inside statistical errors,
indicating that the first term in (8) is the only relevant one
for $E$ near 1 in the infinite volume limit.

The physical picture which emerges from the phase diagram plotted in Fig. 2
is that only one phase does exist in the $(\beta, N)$ plane for massless
compact $QED_3$. This suggests that chiral symmetry be spontaneously
broken for any
number of dynamical fermions. We can not give a definite answer at present
to what extent this
result will remain unchanged when all the higher order
terms in the $N$ expansion of
the effective fermionic action will be included. However we have presented
strong arguments and numerical results which make hard to imagine that
a qualitative change might occur.

This work has been partly supported by MEC (Ministerio de Educacion y Ciencia)
and CICYT (Comision Interministerial de
Ciencia y Tecnologia). X.Q.L. would like to thank the staff of the
Theoretical Physics Department
for helpful discussions and kind assistances.

\eject
%%%%%%%%%%%%%%%%%%%%%%%%%%%%%%%%%%%%%%%%%%

%%%%%%%%%%%%%%%%%%%%%%%%%%%%%%%%%%%%%%%%%%

\eject

\centerline{\bf Figure Captions}

\vskip 0.6cm

Fig. 1. First contribution to the effective fermionic action (8)
versus $E$ at several fermion masses and two
flavors,
obtained through microcanonical simulations.
Statistical errors are invisible at this scale.

\vskip 0.6cm

Fig.2. Phase diagram in the $(\beta, N)$ plane.
Statistical errors are invisible at this scale.


\begin{thebibliography}{9}
\bibitem {15.} {Notice that since the effective fermionic action is non local,
Griffiths demonstration of the convexity of the effective hamiltonian does
not hold.\hfill}
\end{thebibliography}
\end{document}